# Ferroelectricity in Hafnia Controlled *via* Surface Electrochemical State

Kyle P. Kelley,[1,a] Anna N. Morozovska,[2] Eugene A. Eliseev,[3] Yongtao Liu,[1] Shelby S. Fields,[4] Samantha T. Jaszewski,[4] Takanori Mimura,[4] Jon F. Ihlefeld,[4,5] and Sergei V. Kalinin[1]

[1] Center for Nanophase Materials Sciences, Oak Ridge National Laboratory, Oak Ridge, TN 37831, USA

[2] Institute of Physics, National Academy of Sciences of Ukraine, pr. Nauki 46, 03028, Kyiv, Ukraine

[3] Institute for Problems of Materials Science, National Academy of Sciences of Ukraine, Krjijanovskogo 3, 03142, Kyiv, Ukraine

[4] Department of Materials Science and Engineering, University of Virginia, Charlottesville, VA 22904, USA

[5] Charles L. Brown Department of Electrical and Computer Engineering, University of Virginia, Charlottesville, VA 22904, USA

**Abstract:**

Ferroelectricity in binary oxides including hafnia and zirconia have riveted the attention of the scientific community due to highly unconventional physical mechanisms and the potential for integration of these materials into semiconductor workflows. Over the last decade, it has been argued that behaviors such as wake-up phenomena and an extreme sensitivity to electrode and processing conditions suggests that ferroelectricity in these materials is strongly coupled with additional mechanisms, with possible candidates including the ionic subsystem or strain. Here we argue that the properties of these materials emerge due to the interplay between the bulk competition between ferroelectric and structural instabilities, similar to that in classical antiferroelectrics, coupled with non-local screening mediated by the finite density of states at surfaces and internal interfaces. Via decoupling of electrochemical and electrostatic controls realized via environmental and ultra-high vacuum PFM, we show that these materials demonstrate a rich spectrum of ferroic behaviors including partial pressure- and temperature-induced transitions between FE and AFE behaviors. These behaviors are consistent with an antiferroionic model and suggest novel strategies for hafnia-based device optimization.

The discovery of ferroelectricity in hafnia took the scientific community by surprise.[1, 2] These findings represented the first reliable observation of ferroelectricity in binary oxides, suggesting a mechanism very different from that in classical $ABO_3$ perovskites. Similarly, hafnia and related binary oxides allow for integrability in semiconductor manufacturing workflows, a long-standing challenge for the ferroelectric community spurred by potential applications in non-volatile random access memories,[3-6] tunneling barriers,[7-9] field-effect transistors,[10-12] and multiferroic devices.[13, 14] It is notable that initial reports of ferroelectricity in hafnia and zirconia were followed by an exponentially rapid growth of groups active in the area. Furthermore, this discovery has stimulated investigation of ferroelectricity in other binary materials including $Mg_xZn_{1-x}O$, $Al_xB_{1-x}N$, and $Al_{1-x}Sc_xN$ suggesting that ferroelectricity can be ubiquitous in binary nitrides and oxides.[15-21]

From the first reports on ferroelectricity in hafnia, significant controversy emerged on the origins of the switchable spontaneous polarization in this material.[22] Early density-functional theory calculations demonstrated the feasibility of intrinsic polarization instability in these structures;[23-25] however, experimental studies have demonstrated that many aspects of ferroelectric behavior in hafnia were considerably complex compared to classical ferroelectrics. These include a more exaggerated wake-up effect, i.e. when classical ferroelectric hysteresis loops emerge after cycling of the fabricated devices.[26, 27] Similarly, ferroelectric behaviors in hafnia-based systems possess an anomalously strong dependence of electrode interfaces and preparation conditions.[28] Jointly, these observations suggest that ferroelectric-like behavior in these materials either originate or are mediated by alternative mechanisms. The former can include the formation of mesoscale chemical dipoles and vacancy ordering, whereas the latter suggest the potential coupling between ferroelectric and ionic or strained subsystems. Additionally, a large discrepancy between nanoscale free surface and bulk capacitor-based measurements has been observed throughout the community, further suggesting complex mechanisms underpin hafnia's functionality.

Driven by these considerations, a number of groups have proposed that ferroelectricity in hafnia can be mediated by ionic phenomena, most notably oxygen vacancies.[27, 29] Furthermore, the polymorphism of hafnia was posed to be associated with multiple factors including strain, doping, field-induced phase transitions, and changes in oxygen stoichiometry resulting in complexities associated with stabilizing the ferroelectric orthorhombic phase.[30-32] It is worth noting the existing evidence does suggest the ionic subsystem plays a significant role in stabilizing ferroelectricity, but does not unambiguously pinpoint the relevant mechanisms. The reason for this is that many potentially mobile ions such as protons, etc. will not be directly observable via electron microscopy observations. Similarly, the application of an electric field during device testing or *in-situ* experiments perturbs the ferroelectric and ionic subsystems jointly, and does not allow for the decoupling of ferroelectric and ionic effects.

While discussing the state-of-the-art understanding of ferroelectricity in hafnia, we note that over the last decade anomalous ferroelectric-like behaviors were also observed by the classical ferroelectrics community. For example, observations of hysteresis loops and switchable remanent polarization in materials where polarization exists in the bulk but is expected to disappear in thin layers, or in materials that do not have polarization instability were reported.[33] The characteristic aspect of these systems is the presence of a continuum of polarization states (as opposed to binary polarization in classical ferroelectrics), and sensitivity to environmental conditions. Furthermore, often these phenomena could be generally observed on free surfaces via Piezoresponse Force Microscopy (PFM), but not device structures formed from these films.[34] Here, it was argued that the coupling between the bulk ferroelectric instability and surface electrochemistry[35-39] can give rise to new ferroionic states.[40-43] More generally, these studies have pointed to the potentially key role of the potential compensation phenomena at ferroelectric surfaces and interfaces beyond a classical dead-layer model.

Here, we argue that the observed ferroelectric behaviors in hafnia can be ascribed to the interplay between bulk antiferroelectric-instability with non-linear surface (or interface) charge compensation, manifested in the bulk and at the interfaces, respectively. This fundamental disparity in the spatial localization of bulk ferroelectric instability and interfacial behaviors renders the polarization switching and phase evolution non-local, resulting in complex time- and voltage-responses. To explore these assertions, we aim to control the ionic and electrostatic degrees of freedom separately. Note that while this separate control is highly challenging to achieve at the internal interfaces (simply because biasing affects both electrostatic and electrochemical potentials jointly), such control can be readily accomplished at the surfaces via the partial pressure of oxygen and temperature. As such, decoupling of electrochemical and electrostatic controls realized via environmental and ultra-high vacuum PFM, we show that these materials demonstrate a rich spectrum of ferroic behaviors including oxygen partial pressure- and temperature-induced transitions between FE and AFE behaviors.

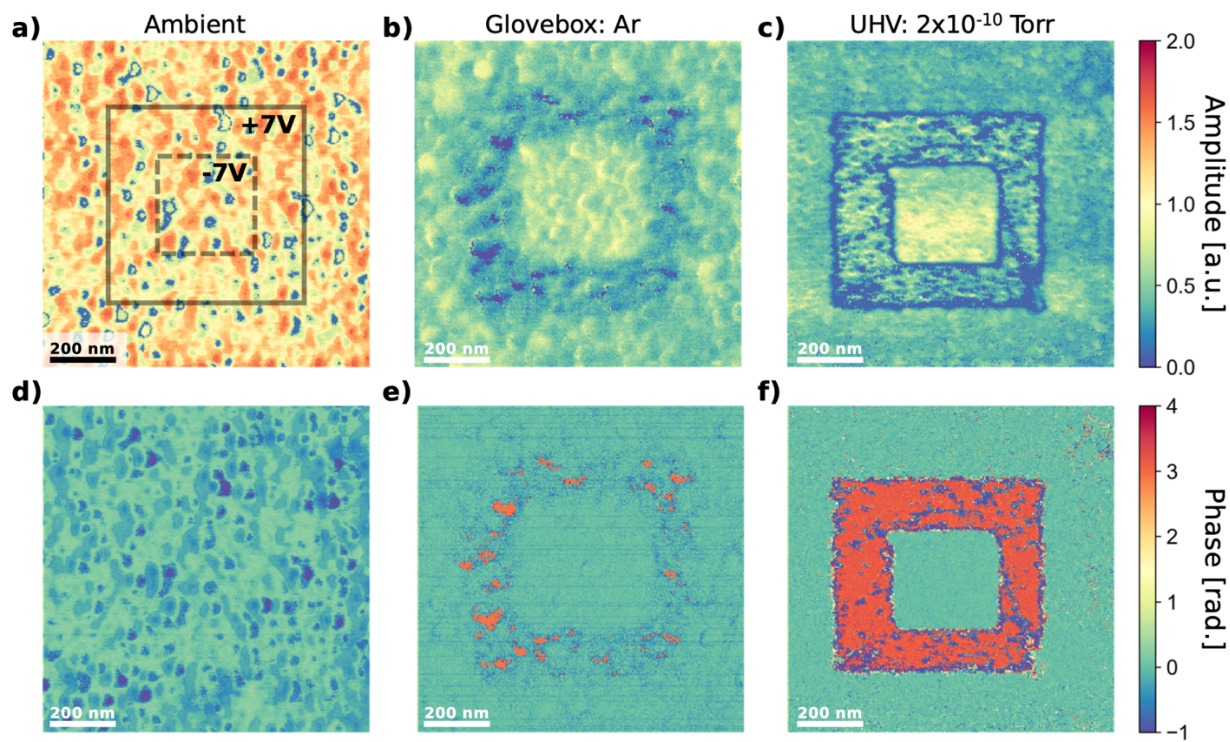

**Figure 1. Band excitation piezoresponse force microscopy ferroelectric poling.** $Hf_{0.5}Zr_{0.5}O_2$ thin films poled via AFM bias of +7 V (600 nm solid box, panel **(a)**) and -7 V (300 nm dotted box, panel **(a)**) in three different environments: **(a,d)** ambient, **(b,e)** argon, and **(c,f)** ultra-high vacuum (amplitude and phase response respectively). Note, images are acquired after poling via bias applied to AFM cantilever.

As a model system to explore hafnia-based antiferroelectric instabilities, we have chosen a 20 nm $Hf_{0.5}Zr_{0.5}O_2$ (HZO) thin film grown via plasma enhanced atomic layer deposition on a silicon substrate with TaN bottom electrodes (see Methods Section). Band excitation (BE) PFM measurements were performed in three different environments, i.e., ambient, glove box (<10 ppm $O_2$) and in UHV vacuum ($2\times10^{-10}$ Torr) to control the effective partial pressure of oxygen. Note that in contrast to single frequency PFM, BE-PFM utilizes a non-sinusoidal signal with a defined band in frequency space to independently detect resonance frequency and response amplitude, ultimately mitigating topographic crosstalk.[44, 45]

Figure 1 shows traditional PFM poling experiments acquired in all three environments using +7 V and -7 V biases to pole 600 nm and 300 nm areas, respectively (solid and dotted boxes,

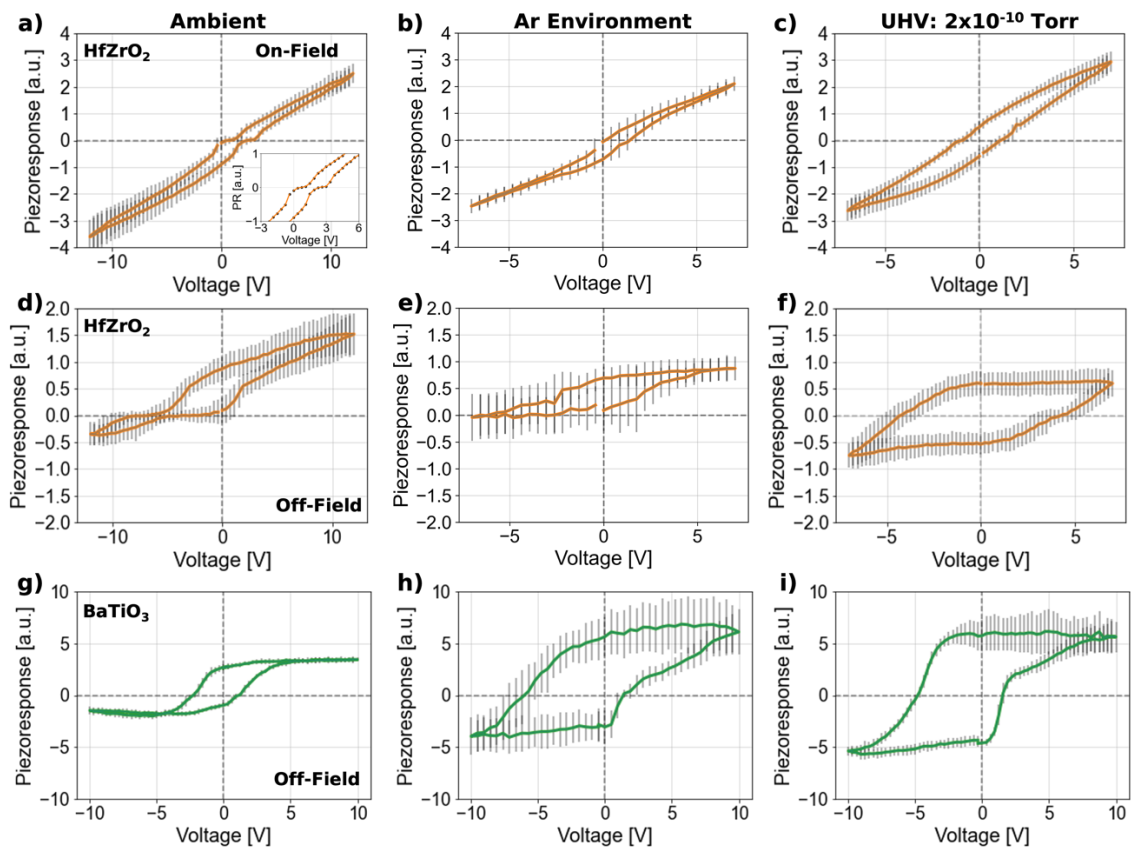

**Figure 2. Band excitation piezoresponse spectroscopy hysteresis loops.** Average on-field and off-field hysteresis loops of Hf$_{0.5}$Zr$_{0.5}$O$_2$ over 2 cycles acquired from a 10x10 grid over a 1μm area in three different environments: **(a,d)** ambient, **(b,e)** argon, and **(c,f)** ultra-high vacuum. For comparison, average off-field hysteresis loops of 80 nm thick BaTiO$_3$ are shown in panels **(g-i)** acquired in ambient, argon, and ultra-high vacuum, respectively. Top **(a-c)** and bottom **(d-i)** panels correspond to on-field and off-field hysteresis loop, respectively. The inset in panel **(a)** shows magnified hysteresis loop where strong pinching is observed. Note, ambient hysteresis loops span ±12V biases, while argon and ultra-high vacuum hysteresis loops span ±7V biases.

Figure 1a). Figure 1a,d displays the BE-PFM amplitude and phase, respectively, acquired after poling in ambient conditions. Here, the formation of domain walls and oppositely poled regions are unobserved. Interestingly, localized regions of the PFM phase (Figure 1d) seem to vary across the region of interest suggesting the presence of additional mechanisms, such as nonuniform polarization or surface states. As the environmental oxygen partial pressure is reduced, i.e. glovebox measurements, localized areas with decreased amplitude are observed (Figure 1b,e) in the +7 V poled region. In contrast, as the environmental oxygen partial pressure is further reduced via UHV conditions, clear domain wall formation and polarization switching is observed within the poled regions (Figure 1c,f) with the presence of localized unswitched areas in the +7 V poled region, suggesting possible localized defect clusters or different crystallographic phases.

Next, HZO hysteresis loops were measured via band excitation piezoresponse spectroscopy (BEPS) in ambient, glove box, and UHV environments at 300 K and the results are shown in Figure 2. Here, classical hysteresis loops measured via PFM experiments are presented in the off-field state, where the PFM signal is measured at the zero-bias state after the application of a short bias pulse. This experimental protocol is chosen to minimize the contribution of electrostatic forces in the measured signal.[46, 47] In the interpretation of these measurements, it is postulated that the domain formed below the tip does not appreciably relax on the transition to zero probe bias, i.e. the domain walls are strongly pinned.[48] However, this assumption is not necessarily justified for materials, such as ferroelectric relaxors with highly mobile domain walls, or materials with competing order parameters such as antiferroelectrics. Hence, here we present both on-field and off-field measurements.

Figure 2a,d shows the on-field and off-field switching behavior in ambient conditions, respectively. Remarkably, spectroscopy biases up to 12 V can be applied to the HZO films without dielectric breakdown. Here, clear pinching of the on-field hysteresis loop close to 0 V is observed, indicative of antiferroelectric-like behavior. Notably, the pinched loop is offset from 0 V suggesting the presence of a built-in potential (Figure 2a inset). Examination of the off-field loop also shows antiferroelectric-like behavior, suggesting metastability of the bias-induced polar phase. However, in an argon environment, i.e. glovebox measurements, a lower degree of pinching accompanied with a narrowing of the hysteresis loop is observed in the on-field state with no remanent ferroelectric switching observed in the off-field state (Figure 2b,e), indicating a paraelectric-like phase. In contrast, in UHV conditions ($2 \times 10^{-10}$ Torr) a widening of the on-field loop with the absence of pinching is observed, indicating the onset of the ferroelectric state (Figure 2c). Moreover, the off-field hysteresis loops show clear remanent switching (Figure 2f).

It is important to note, a similar trend can be observed when cycling $Hf_{0.5}Zr_{0.5}O_2$ within a capacitor geometry (Supplementary Figure S8) where cycling facilitates a more traditional ferroelectric response (i.e. wake-up phenomena). In particular, scanning probe measurements performed in a reducing environment (i.e. in UHV) on a free surface produce similar ferroelectric responses as bulk measurements (i.e. capacitor geometries) post cycling, suggesting a common mechanism.

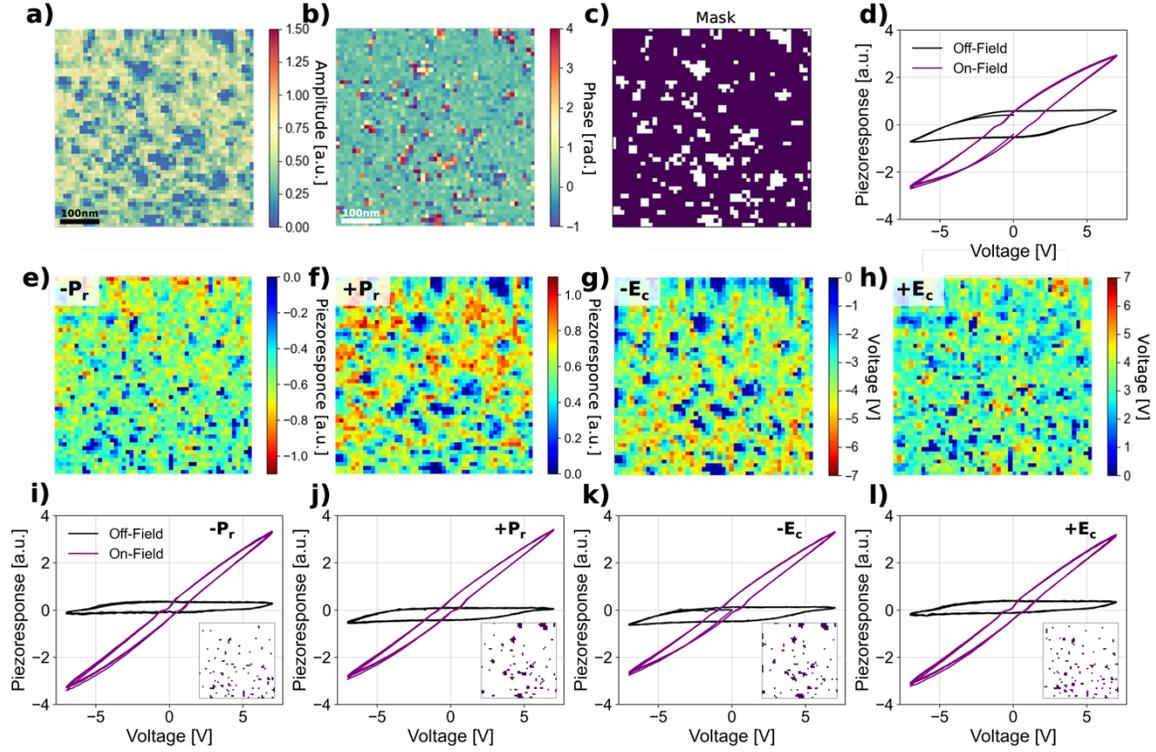

**Figure 3. High density band excitation piezoresponse spectroscopy in ultra-high vacuum.** BE piezoresponse spectroscopy off-field **(a)** amplitude and **(b)** phase of 50x50 high density spectroscopy grid acquired over 500 nm at $V, t = 0$ (V,s). **(c)** Masked off-field amplitude using a threshold of 0.2 to identify regions of higher response and **(d)** corresponding average on-field and off-field (purple, black respectively) hysteresis loops. **(e,f)** Remanent polarization and **(g,h)** coercive field maps derived from high density spectroscopy grid in the off-field state. **(i-l)** Average loops acquired from masked regions determined from thresholding remanent polarization and coercive field maps, i.e. low $P_r$ and $E_c$ regions.

In light of these observations, it is important to consider how traditional ferroelectrics, e.g. $BaTiO_3$, behave in all three environments. As such, for comparison, average off-field hysteresis loops of 80 nm thick $BaTiO_3$ are shown in Figure 2g-i. Here, clear hysteretic switching can be observed in all environments, as expected; however, differences in the hysteresis loop shape are readily observed indicating environmentally induced surface effects, such as the propensity for surface adsorbates or oxygen vacancy formation at the surface, can have substantial impact on the ferroelectric behavior.[49]

To gain further insight into the localized ferroelectric switching dynamics in HZO observed under low environmental oxygen partial pressure, we perform a high density 50x50 spectroscopy grid (Figure 3) over a 500 nm region in UHV, approaching the resolution limit of PFM. Figure 3a,b shows the amplitude and phase at $t=0$ (0V) where clear inhomogeneous response can be observed. Here, we aim to evaluate the hysteresis loops in selective regions to deconvolute the spatially inhomogeneous response. As such, we apply a threshold mask (amplitudes greater than 0.2) to Figure 3a to isolate the higher response regions, shown as the masked purple regions in Figure 3c. The average on-field and off-field response from the masked region is shown in Figure 3d. Importantly, the masked regions show a similar response to Figure 2c,f (i.e. ferroelectric switching).

Next, we calculate remanent polarization, $P_r$, and coercive field, $E_c$, maps from the dense spectroscopy grid (in the off-field state) determined by the zero bias and zero polarization crossings, respectively, yielding four maps: $-P_r$, $+P_r$, $-E_c$, and $+E_c$ (Figure 3e-h). Qualitatively, the regions of low remanent polarization and coercive field are in good agreement with the initial low response regions observed in Figure 3a. To identify the localized switching behavior of these regions, i.e. blue clusters in Figure 3e-h, we threshold the remanent polarization and coercive field maps to isolate the regions (similar to the methodology employed in Figure 3c) and plot the corresponding average hysteresis loops (see Methods Section). Interestingly, regions of low $-P_r$ and $+E_c$, and low $+P_r$ and $-E_c$ are spatially correlated as seen in Figure 3e,h and Figure 3f,g respectively. Furthermore, the average on-field and off-field hysteresis loops derived from these regions are strikingly similar. Specifically, for regions of low $-P_r/+E_c$, the average on-field loop has a high degree of pinching isolated to the negative branch of the loop, while the average off-field loop shows mostly positive piezoresponse. In contrast, in the regions of low $+P_r/-E_c$, the average on-field loop has a high degree of pinching isolated to the positive branch of the loop, while the average off-field loop exhibits mostly negative piezoresponse. Here, we posit at 300 K in UHV, the HZO thin film is on the boundary of antiferroelectric instability where the observed spatial inhomogeneities suggest the presence of localized antiferroic states, or crystallographic phases (e.g., tetragonal phase).

These observations suggest the electrochemical state of the surface strongly couples to the ferroic behavior in the hafnia system. In other words, as the environmental oxygen partial pressure is reduced, the surface charge is expected to increase due to the chemical potential driving force which both reduces compensating mechanisms (e.g. surface absorbates or ionic species) and the oxygen content at the surface, potentially giving rise to the observed ferroelectric stability. As such, it is worth mentioning the placement of conductive electrodes on hafnia's surface can also provide the necessary interfacial charge to stabilize the ferroelectric behavior, congruent with previous reports.[50]

To gain further insight, we apply a previously developed formalism for an antiferroelectric film coupled to surface electrochemical reactions.[51] This phenomenological formalism considers competing polar and structural anti-polar distortive long-range orders in (Hf,Zr)O$_2$ at the mesoscopic level with the possible role of doping and oxygen vacancies, which could contribute to the origin of hafnia's ferroelectric properties at the microscopic level and can be involved in the framework of mean-field approach.[52]

Briefly, we consider a material with the competing incipient ferroelectric and structural instabilities, for which the simplified bulk free energy is an expansion on even powers of the polar ($P_i$) and structural anti-polar ($A_i$) order parameters:

$$F_{bulk} = \int_{V_f} d^3r \left( \frac{a_i}{2} P_i^2 + \frac{a_{ij}}{4} P_i^2 P_j^2 - P_i E_i + \frac{\chi_{ij}}{2} P_i^2 A_j^2 + \frac{b_i}{2} A_i^2 + \frac{b_{ij}}{4} A_i^2 A_j^2 \right), \quad (1)$$

Here $V_f$ is the film volume. The coefficients $a_i$ and $b_i$ linearly depend on temperature $T$; in particular $a_i = \alpha_{TP}(T_P - T)$ and $b_i = \alpha_{TA}(T_A - T)$. They change sign at the virtual Curie temperature $T_P$ and AFE temperature $T_A$, respectively, with the constants $\alpha_{TP} > 0$ and $\alpha_{TA} > 0$. All other tensors in Eq. (1) are regarded as temperature-independent. The tensors $a_{ij}$ and $b_{ij}$, are positively defined for the stability of functional $F_{bulk}$. Tensor $\chi_{ij}$ is responsible for the coupling between the order parameters, and the sign and values of its components are such that the material is structurally distorted and non-polar ($A \neq 0, P = 0$) in the ground state.

The application of an electric field can give rise to the polar phases $P \neq 0$, with or without the structural distortion. Note that transitions from the structural phase ($A \neq 0, P = 0$) to the polar phase ($A = 0, P \neq 0$) defines antiferroelectric properties of the material.[53, 54] Here, the electric field $E_i$ is related with electric potential $\phi$ in a conventional way, $E_i = -\frac{\partial \phi}{\partial x_i}$, and the potential is defined as the solution of electrostatic equations with corresponding boundary conditions (see Supplementary Information for details). For classical ferroelectric boundary conditions, i.e., a defined potential at the electrodes in the presence of the dielectric dead layer, the evolution of the system is well known.[55] However, the situation changes drastically in the presence of finite surface (or interface) density of ionic and electronic states. Here we consider the classical Langmuir adsorption behavior as originally proposed by Stephenson and Highland,[37-39] and a linear relaxation model to describe the temporal dynamics of the positive ($\sigma_p$) and negative ($\sigma_n$) surface charge densities:

$$\tau_p \frac{\partial \sigma_p}{\partial t} + \sigma_p = \sigma_{p0}[\phi], \quad \tau_n \frac{\partial \sigma_n}{\partial t} + \sigma_n = \sigma_{n0}[\phi], \quad (2)$$

where $\tau_p$ and $\tau_n$ are relaxation times for the ionic and electronic charges, respectively. In the forthcoming analysis, we assume that electronic relaxation times are much smaller than those for the ionic species. The dependence of equilibrium charge densities $\sigma_{p0}[\phi]$ and $\sigma_{n0}[\phi]$ on the electric potential $\phi$ is controlled by the concentration of surface ions at the interface $z = 0$, as proposed by Stephenson and Highland:[37-39]

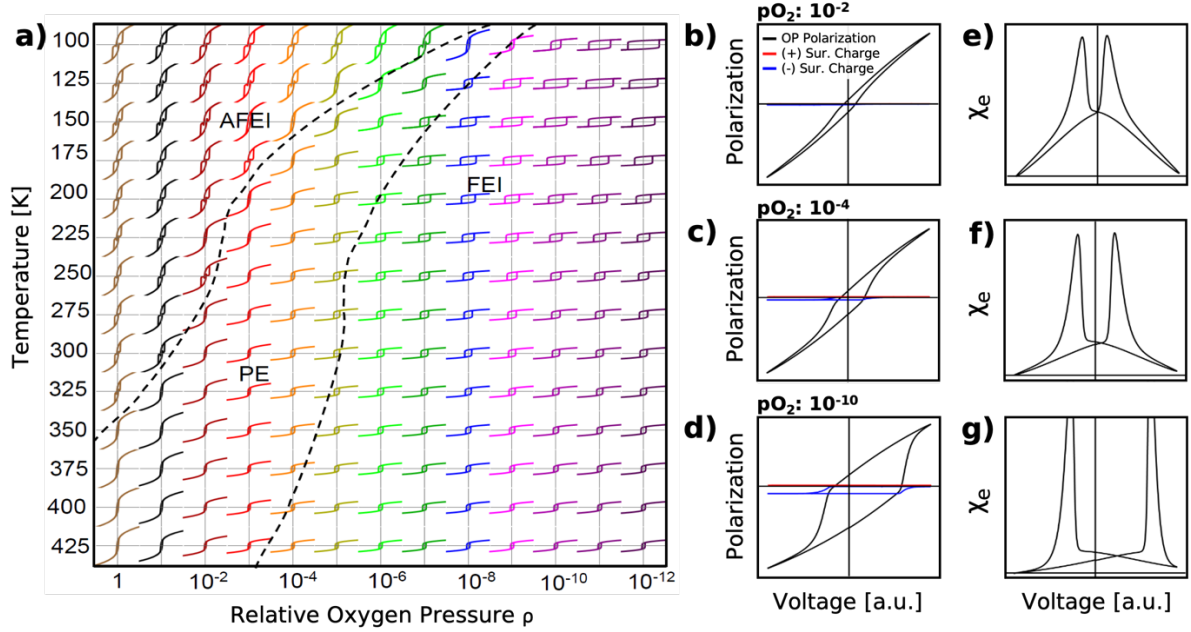

**Figure 4. (a)** Phenomenological Landau-type modeling of HZO temperature versus relative oxygen pressure phase diagram. The thickness of HZO film is $h = 20$ nm. Dashed curves separate different type of loops corresponding to the nonpolar AFEI-phase, the intermediate PE-like phase and the FE-like FEI phase. **(b)-(d)** The dynamic voltage dependencies of the out-of-plane polarization (black curves), positive (red curves) and negatives (blue curves) surface charges calculated for decreasing oxygen pressure ($10^{-2}$, $10^{-4}$ and $10^{-10}$ $pO_2$) and different relaxation times of the screening charges, $\tau_n \ll \tau_p$. **(e)-(f)** Corresponding dependences of the dielectric susceptibility ($\chi_e$). Parameters used in calculations are listed in **Table S1.** Note that plots (b)-(g) are calculated for higher frequency of applied bias than the quasi-static polarization loops shown in the diagram (a). Therefore, the quasi-static and dynamic loops look different.

$$\sigma_{0i}[\phi] = \frac{eZ_i}{N_i}\left(1 + \rho^{-1/n_i} \exp\left(\frac{\Delta G_i^{00} + eZ_i\phi}{k_B T}\right)\right)^{-1}, \quad (3)$$

where $e$ is an elementary charge, $Z_i$ is the ionization degree of the surface ions/electrons, $1/N_i$ are saturation densities of the surface ions. A subscript $i$ designates the summation on positive ($i = p$) and negative ($i = n$) charges, respectively; the dimensionless ratio $\rho = \frac{p_{O2}^{00}}{p_{O2}}$ is the relative partial pressure of oxygen (or other ambient gas), $n_i$ is the number of surface ions created per gas molecule. The overall behavior of the system has been explored theoretically in the previous work[51] and in the Supplementary Information.

Here, we use this formalism to plot the phase diagram of a 20-nm thick HZO film as temperature versus relative oxygen pressure (Figure 4a) where the dashed curves separate different type of loops corresponding to the nonpolar antiferroelectric-like ferroionic phase (AFEI), the intermediate paraelectric-like phase (PE), and the ferroelectric-like ferroionic phase (FEI). This classification is valid for quasi-static changes of the applied voltage and is based on the shape of polarization hysteresis loops, which continuously transforms from a double AFE-type loop through hysteresis-less PE curve towards a single FE-type loop as shown in Figure 4a. Importantly, the experimentally measured hysteresis loops acquired as a function of relative oxygen partial pressure are in great qualitative agreement with the phenomenological Landau-type modeled loops, suggesting the experimentally observed behavior is due to the proposed AFEI-PE-FEI transition mechanisms, i.e., the interplay between bulk antiferroelectric-instability with non-linear surface charge compensation. Also, it is important to consider that the modeled behavior indicates the AFE-type loops transform to FE-type with the increase of the applied voltage frequency (Figures S1-S7 in the Supplement Information), which is potentially important for hafnia-based applications.

A critical component of the modeled behavior is the dynamic voltage dependencies on the out-of-plane polarization (black curves), and positive (red curves) and negative (blue curves) surface charges, which are calculated for decreasing oxygen pressure and different relaxation times between the positive and negative screening charges, $\tau_n \ll \tau_p$ (Figure 4b-d). A clear increase in hysteresis loop opening, or ferroelectric-like behavior, can be seen with decreasing relative oxygen partial pressure (Figure 4b-d), which is accompanied by an increase in negative surface charges (potentially oxygen vacancies) and relatively constant positive surface charges. Congruently, the corresponding dielectric susceptibilities $\chi_e$ are shown in Figure 4e-f (calculation parameters selected based on experimental observations are listed in Table S1 in Supplementary Information), which have two peaks corresponding to FE-type states. However, the "base" of the peaks reflects one more complex feature of the loops inherent to FEI states. Note that the features characteristic to AFEI states can be seen at much lower frequencies of applied bias, and they represent the continuous transformation of the "base" into two additional peaks (see Supplementary Information).

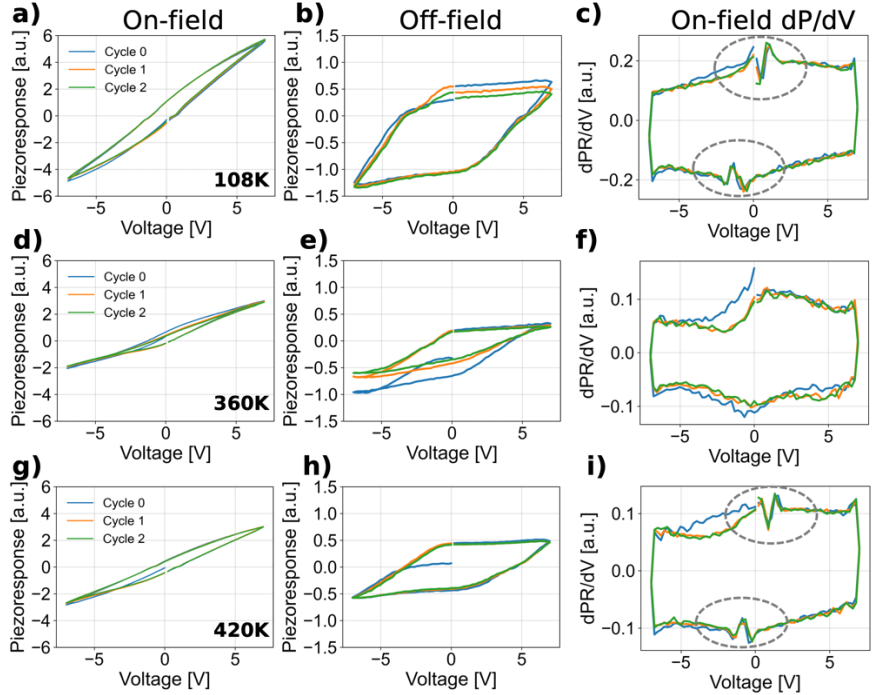

**Figure 5. Temperature dependent band excitation spectroscopy in ultra-high vacuum.** Average band excitation piezoresponse spectroscopy hysteresis loops acquired over a 10x10 grid, 1μm area, and 3 cycles. Measurements acquired at **(a-c)** 108K, **(d-f)** 360K, and **(g-i)** 420K. Note, columns represent **(a,d,g)** on-field, **(b,e,h)** off-field, and **(c,f,i)** differential on-field dP/dV hysteresis loops.

We further proceed to explore temperature dynamics in an effort to probe the complex HZO ionic-electrostatic mediated phase diagram. Here, we note that in classical ferroelectrics the hysteresis loops are weakly dependent on temperature away from the phase transition, and generally expected to remain symmetric. At the same time, ferroionic systems were predicted to give rise to a broad range of temperature-mediated behaviors, including the transitions between ferroelectric-like and antiferroelectric like behaviors.

Here, we collected BEPS hysteresis loops over a 108 to 420 K temperature range in UHV (Figure 5) to vertically transverse the predicted complex phase diagram in a low oxygen partial pressure environment. Figure 5a-i shows average on-field, off-field, and on-field differential hysteresis loops acquired at 108, 360, and 420 K from a 1 μm area (10x10 grid). At 108 K, the on-field loops (Figure 5a) display an observable pinching near 0 V while the off-field (Figure 5b) loops have an asymmetric response, particularly along the piezoresponse axis, similar to Figure 3i-l. Interestingly, when evaluating the on-field differential d$P$/d$V$ (Figure 5c), two clear kinks are observed corresponding to pinching of the on-field loop (dotted circles Figure 5c,i). Here, we identify the first order changes in d$P$/d$V$ to be associated with antiferroelectric behavior. Upon increasing the temperature to 360 K, no pinching was observed, which is readily apparent from the d$P$/d$V$ loops (Figure 5f). Furthermore, the off-field loops at 360 K have distinctly different switching behavior; namely, a decrease in the coercive field relative to hysteresis loops acquired at 108 K, which is in remarkably good agreement with the predicted behavior shown in Figure 4a. Lastly, at 420 K, the on-field loop pinching (Figure 5g) reappears, which is confirmed via kinks in the d$P$/d$V$ spectra (Figure 5i), suggesting a temperature-induced ferroelectric-antiferroelectric instability. Exploring the off-field response at 420 K reveals symmetric hysteresis loops with a larger coercive voltage, potentially originating from mechanisms such as ionic motion, which is certainly plausible at elevated temperatures. Overall, the temperature dependent BEPS measurements are in qualitative agreement with the modeled results with the exception of observations at higher temperatures, i.e. at 420 K antiferroelectric-like behavior is observed suggesting the presence of additional coupled mechanics (e.g. ionic motion).

To summarize, here we reveal strong dependence of the ferroelectricity in hafnia on the external environment, explored over a range of partial pressures and temperatures. These observations confirm that electrochemical state of the surface strongly couples to the ferroelectric phase transition and stability in these materials and is inseparable from ferroelectricity. These observations can be described in terms of the antiferroionic model of coupled antiferroelectric bulk and surface electrochemical phenomena. Furthermore, we explored the key role of atmosphere in changing the charge state of the surface, filling the ionic density of states. Here, similar mechanisms are expected to be valid for internal interfaces that can trap ionic and electronic charges. These mechanisms can provide a general explanation for the range of the unusual phenomena obtained in hafnia and similar binary oxides, including slow evolution of the ferroelectric properties, dependence on processing conditions, and the observed discrepancy between ferroelectric measurements acquired on a free surface and in a capacitor geometry. This, in turn, creates novel opportunities for the discovery and optimization of these material systems, enabled via independent ionic control.

**Acknowledgements**:

This effort (K.P.K., Y.L., S.S.F., S.T.J., T.M., J.F.I., and S.V.K.) was supported as part of the center for 3D Ferroelectric Microelectronics (3DFeM), an Energy Frontier Research Center funded by the U.S. Department of Energy (DOE), Office of Science, Basic Energy Sciences under Award Number DE-SC0021118. The scanning probe microscopy research was performed and partially supported at Oak Ridge National Laboratory's Center for Nanophase Materials Sciences (CNMS), a U.S. Department of Energy, Office of Science User Facility. S.T.J. acknowledges support from the U.S. National Science Foundation's Graduate Research Fellowship Program via grant number DGE-1842490. A.N.M. was supported by the National Academy of Sciences of Ukraine (the Target Program of Basic Research of the National Academy of Sciences of Ukraine "Prospective basic research and innovative development of nanomaterials and nanotechnologies for 2020 - 2024", Project № 1/20-Н, state registration number: 0120U102306) and received funding from the European Union's Horizon 2020 research and innovation programme under the Marie Skłodowska-Curie grant agreement No 778070.

**Materials and methods**

*Material Synthesis*

Metal-Insulator-Metal (MIM) TaN/HZO/TaN devices were prepared on 500 μm thick silicon substrates. The planar bottom TaN electrode was deposited through a DC sputtering process from a sintered TaN target to a thickness of 100 nm at room temperature. Next, an Oxford FlexAL II Plasma-Enhanced Atomic Layer Deposition system was utilized to grow a $Hf_{0.5}Zr_{0.5}O_2$ film to a thickness of 20 nm at 260 °C using tetrakis(ethylmethylamido)hafnium (TEMA Hf) and tetrakis(ethylmethylamido)zirconium (TEMA Zr) as $HfO_2$ and $ZrO_2$ precursors, respectively, in 5:5 super cycle ratios to control film composition. An oxygen plasma was employed as the oxidant.

Following HZO deposition, a 20 nm thick planar TaN top electrode was prepared through an identical process to the bottom electrode, and the film stack was annealed within an Allwin21 AccuThermo 610 Rapid Thermal Processor for 30 seconds under a $N_2$ atmosphere. For the sample for piezoresponse force microscopy measurements, the entire film was exposed to a SC-1 (5:1:1 $H_2O$:30% $H_2O_2$ in $H_2O$:30% $NH_4OH$ in $H_2O$) etch solution at 60 °C for 45 minutes to fully remove the top electrode layer. To prepare electrodes, 50 nm thick Pd electrodes were deposited through a shadow mask and the exposed TaN was removed using the SC-1 treatment. The discrete Pd pads served as a hard mask.

*Polarization Measurements*

Polarization versus electric field measurements were performed with a Radiant Technologies Precision LC II ferroelectric tester with a measurement period of 1 ms. An Instec hot-stage probe station equipped with a turbo molecular pump was used to measure performance at in ambient and high vacuum conditions at 300 K, 360 K, and 420 K.

*Atomic Force Microscopy Measurements*

All piezoresponse force microscopy measurements were captured using Budget Sensor Multi75E-G Cr/Pt coated AFM probes (~3 N/m). Band excitation was achieved by coupling the AFMs with an arbitrary wave generator and data acquisition electronics based on a National Instruments fast DAQ card. Custom software was used to generate the probing signal and store local BE and hysteresis loops. All band excitation PFM measurements were collected using frequencies ranging from 300-400 kHz, with subsequent simple harmonic oscillator fits applied to the collected spectra to extract amplitude and phase at the resonance frequency. Ambient PFM measurements were taken using an Oxford Instruments Cypher atomic force microscope. Glovebox measurements were taken using a Bruker Dimension Icon fitted inside a MBRAUN MB200B series glovebox. Ultra-high vacuum measurements were deployed via a modified ultra-high vacuum Omicron AFM-STM microscope controlled from a Nanonis real time controller at a pressure of ~$2 \times 10^{-10}$ Torr.

*Theoretical Modeling and Analysis*

Data supporting theoretical results was visualized in Mathematica 12.2 [https://www.wolfram.com/mathematica] and can be found at the Notebook Archive (https://notebookarchive.org/xxxx). Thresholding for coercive field and remanent polarization maps were acquired using a ±1V and ±0.2 a.u. threshold, respectively.